# Scalable and deterministic construction of moiré superlattice in 2D materials using stressor films

Yu-Mi Wu[1], Sihun Lee[1], Yufeng Xi[2], Stephen D. Funni[1], Saif Siddique[1,3], Natalie L. Williams[4], Giovanni Sartorello[5], Hesam Askari[2], and Judy J. Cha[1,5a]

[1]*Department of Materials Science and Engineering,*

*Cornell University, Ithaca, NY 14853, USA*

[2]*Department of Mechanical Engineering,*

*University of Rochester, Rochester, NY 14611, USA*

[3]*Kavli institute at Cornell for Nanoscale Science,*

*Ithaca, NY 14853, USA*

[4]*Department of Chemistry and Chemical Biology,*

*Cornell University, Ithaca, NY 14853, USA*

[5]*Cornell Nanoscale Facility, Cornell University, Ithaca, NY 14853, USA*

## Abstract

Moiré superlattice in two-dimensional (2D) materials provides a powerful platform to engineer emergent electronic states, yet the construction of moiré superlattices remains lab-scale, involving much trial and error and with little control. Here, we demonstrate the construction of a heterostrain-induced moiré superlattice in transition metal dichalcogenides using a scalable process that deterministically induces strain to 2D materials. By applying patterned thin-film stressors and probing the resulting structures with scanning transmission electron microscopy, we directly resolve the induced heterostrain, lattice deformations, and stacking variations that produce the moiré superlattice. We find that uniaxial and biaxial heterostrain give rise to distinct moiré patterns, including stripes and distorted hexagonal patterns. With this approach, we create in-plane polar distortions and thus in-plane polarization at the domain boundaries of the moiré superlattice in $MoS_2$. The deterministic and scalable construction of moiré patterns using a well-established scalable process opens opportunities to design new moiré geometries in 2D materials.

Corresponding Author: [a] jc476@cornell.edu

Moiré superlattices form when layered materials are stacked with a small relative twist or lattice mismatch, producing periodic modulations on a length scale much larger than the atomic lattice [1–3]. They have led to many ground-breaking phenomena including superconductivity [4, 5], topological edge states [6, 7], exciton trapping [8], correlated insulating phases [9, 10], and spontaneous polarization [11–13]. At the microscopic level, local stacking orders in the moiré superlattice control the interlayer coupling strength, intralayer strain and lattice reconstruction, and form moiré domains whose boundaries play a central role in determining the electronic ground states [14–16]. Thus, controlling the layer stacking and strain has become a key strategy to tailor the properties of moiré superlattices.

To date, most studies of moiré systems have focused on twist-controlled bilayers such as twisted bilayer graphene [4, 5, 9, 17] and twisted transition metal dichalcogenides (TMDs) such as bilayer $WSe_2$ [18, 19], $WS_2$ [20, 21], $WTe_2$ [22, 23] and $MoTe_2$ [24–26]. Strain-induced moiré patterns have been realized in vertically stacked heterobilayers with intrinsic lattice mismatch, such as $WSe_2/WS_2$ [27], $MoTe_2/WSe_2$ [28, 29], and graphene/hexagonal boron nitride [30, 31], where the difference in lattice constants produces moiré modulations. More recently, attention has turned to heterostrain in homostacks [32], a state in which adjacent layers of the same material experience different lattice deformations, offering an additional means to further tune moiré geometry and electronic structure. While its potential has been explored in theory [33–37], experimental progress remains limited. Most observations arise from unintentional strain, and only a few device demonstrations have achieved controlled heterostrain [38–40]. Most importantly, both twist-controlled and strain-controlled moiré constructions remain lab-scale, without obvious pathways for scalability and industry adoption.

Strain engineering via stressor film deposition is a standard industry process in complementary metal-oxide-semiconductor (CMOS) integrated circuit fabrication, where controlled stress is engineered to enhance carrier mobility [41, 42]. Adapted to 2D materials, stressor thin films can transfer strain to the underlying lattice, providing a scalable and controllable route to achieve heterostrain [43, 44]. By further patterning the stressor films with lithography, both biaxial and uniaxial strains can be locally introduced in a site-specific manner [45, 46]. The strain orientation can be aligned with specific crystallographic directions, while its spatial distribution and magnitude can be tuned by adjusting the geometry and thickness of the stressor films. This tunability enables precise control of strain fields in the materials.

Building on this capability, we report the construction of heterostrain-induced moiré patterns in MoS$_2$. Using lithographically patterned stressor thin films, we induce controlled heterostrain and directly visualize the resulting moiré superlattices with atomic-resolution scanning transmission electron microscopy (STEM), four-dimensional STEM (4D-STEM), and theoretical calculations. We show that the patterned stressor films produce both uniaxial and biaxial heterostrain in MoS$_2$, giving rise to stripe- and hexagon-like moiré arrangements. Using this approach, we induce in-plane polarization in non-polar MoS$_2$. Detailed analysis of the local lattice structure reveals the direct correlation between the lattice reconstruction and induced polarization. We thus demonstrate a scalable and deterministic construction of moiré superlattices in 2D materials, that is readily adoptable in semiconductor industry.

**RESULTS**

**Patterned stressor films and heterostrain**

Experiments were conducted on MoS$_2$ flakes exfoliated from single crystals of 2H-MoS$_2$. We first transferred MoS$_2$ flakes onto Si substrates using Scotch tape exfoliation and a polypropylene carbonate (PPC) stamp. The flakes were then transferred onto thin SiN$_x$ membrane TEM grids for high-resolution STEM imaging. Next, a patterned stressor film was deposited onto exfoliated MoS$_2$ flakes via e-beam lithography and e-beam evaporation (see Methods and Supplementary Fig. 1 for details). The compressive stressor film consists of a 50 nm thick SiO$_2$ layer sandwiched between two 10 nm thick Al$_2$O$_3$ layers and was patterned into 10 $\mu$m-wide stripes spaced 10 $\mu$m apart. The resulting sample geometry is MoS$_2$ flakes capped with a patterned stressor film, as illustrated in Fig. 1a.

To confirm the presence of strain in MoS$_2$ flakes due to the stressor film, we first performed Raman spectroscopy since the $E_{2g}$ mode is highly sensitive to in-plane strain as it reflects in-plane intralayer vibrations [47, 48]. Figure 1b shows an optical image of a ~20 nm-thick MoS$_2$ flake covered with the patterned stressor film on a Si substrate for Raman measurements. We focus on the $E_{2g}$ mode at 384 cm$^{-1}$ across regions from the uncapped to the stressor-capped areas of the flake (Fig. 1c). The $E_{2g}$ mode first shifts to higher frequencies near the stressor edge, indicative of an in-plane compressive strain. Within the capped region under the stressor, the $E_{2g}$ mode red-shifts, indicating tensile strain transferred by the stressed film. The change of the peak position of the $E_{2g}$ mode is shown in Fig. 1e. The out-of-plane vibrational mode $A_{1g}$ at 409 cm$^{-1}$ also shows a downshift in the capped region (Fig. 1d,e). The $A_{1g}$ peak is sensitive to carrier concentration [49], and the unchanged full width at half maximum (FWHM) of the $A_{1g}$ peak indicates that no additional defects were introduced during film evaporation (Supplementary Fig. 2). This suggests that the $A_{1g}$ peak shift is from an accompanying out-of-plane strain in the capped region [48]. The observed Raman spectral shifts due to stressor films are consistent with previous reports [45,50], confirming that the patterned stressor film effectively induces strain in MoS$_2$ (Supplementary Fig. 3). Microscopically, the stressor film edge exerts a lateral force that pushes the underlying flake sideways, inducing a localized compressive strain near the edge and tensile strain under the stressor film (Fig. 1a schematic).

Next, we investigate how deeply the applied strain penetrates into the MoS$_2$. Cross-sectional STEM imaging directly resolves in-plane lattice distortions layer-by-layer with atomic resolution. Figure 1f shows the atomic resolution high-angle annular dark field (HAADF)-STEM image of the top several MoS$_2$ layers viewed along the [$\bar{1}$100] direction near the edge of the stressor film (see Supplementary Fig. 4). In the 2H stacking polytype (space group *P*63/*mmc*), Mo atoms in every second layer are vertically aligned. However, a closer look at the topmost layer reveals a gradual shift of Mo positions: the left side retains the 2H registry, while on the right, the Mo atoms in the first and third layer are no longer aligned. This indicates a local stacking rearrangement driven by the stressor-induced strain in the topmost MoS$_2$ layer (Supplementary Fig. 4). Figure 1g shows the averaged in-plane Mo-Mo distance for each layer measured by quantitative atom tracking, which shows that the topmost layer is under compressive strain of ~1 %, and the second layer is under ~0.1% strain, while the rest of the layers remain unstrained.

To confirm these observations, Molecular Dynamics (MD) simulations were performed on a five-layer MoS$_2$ model using the LAMMPS package [51] with the REBOMoS potential [52] for intralayer interactions and a Lennard-Jones potential for interlayer coupling (see Methods for details) [37]. After energy minimization, a total of 1% compressive strain was applied to the top layer through incremental ramping while the bottom layer was fixed to mimic the substrate constraint. The simulated Mo-Mo distances (Fig. 1g) agree well with the experimentally measured values. The simulated atomic configuration and magnified stacking views (Fig. 1h) also match the stacking rearrangement observed in STEM. In aggregate, the experimental and simulated results show that the strain imparted by the patterned stressor film is mainly concentrated in the top two layers of MoS$_2$.

**Moiré pattern formation by heterostrain**

Having identified the heterostrain in the top two layers of MoS$_2$, we now focus on plan-view imaging to visualize the structural arrangements at the nanoscale. Figure 2a shows an optical image of a MoS$_2$ flake capped with a ∼10 $\mu$m-wide stressor film on a holey SiN$_x$ membrane. To measure the local strain, we use a nanobeam 4D-STEM approach [53] combined with exit-wave power cepstrum (EWPC) analysis [54], which enables high precision strain mapping across large areas ranging from a few to several hundred nanometers while being robust against sample tilts and noise in electron diffraction patterns. Figure 2b shows a reconstructed, virtual annular dark-field STEM image from the 4D-STEM data, focusing on the region near the stressor film edge where the strain variations are most pronounced. A moiré superlattice with varying domain size and patterns is clearly visible. The corresponding strain maps (Fig. 2c-e) reveal a largely biaxial tensile strain of ∼0.3% in MoS$_2$ along with shear components under the stressor and near the stressor edge. We note that the measured strain values represent the projected strain averaged through the entire ∼20 nm flake thickness, while in reality the top two layers experience the strain of ∼1% and 0.1%, as shown in Fig. 1f,g.

Regions away from the stressor film exhibit a predominantly uniaxial compressive strain. This transition from a biaxial strain near the stressor to uniaxial strain away from the stressor produces a distinct change in the resulting moiré patterns. Directly under the stressor film (top part of Fig. 2b), we observe distorted hexagonal (almost diamond-like) domains that are elongated perpendicular to the stressor edge, with domain boundaries appearing as bright lines. The larger strain in this region results in denser moiré domains with moiré periodicity of 30-40 nm. Moving farther from the stressor, the domains become progressively larger and distorted, eventually evolving into a stripe pattern under uniaxial compressive strain away from the stressor film. These moiré patterns persist over several micrometers. We note that defects introduced during the exfoliation locally modulate the strain field and introduce discontinuities in the moiré periodicity, particularly the diagonal straight fold in Fig. 2b. Nevertheless, on the right side of the fold the moiré pattern evolves from hexagonal to stripe, and merges into the stripe domains at the bottom of the flake. Strain maps further show that the fold perturbs the lattice only along the narrow line, and the dominant variations align parallel to the stressor edge across both sides of the fold. The overall resulting moiré orientation remains perpendicular to the stressor edge. As a control, pristine flakes without stressor films show no moiré patterns or evident bending or defects (Supplementary Fig. 5), confirming that the observed moiré structures are induced by the stressor. Stressor-induced moiré patterns are also observed in rhombohedral-stacked 3R-MoS$_2$ flakes (Supplementary Fig.6), showing that the effect is not limited to the 2H phase.

**Atomic structure of reconstructed lattice**

The stressor film-induced strains thus alter the layer stacking in the top few layers of MoS$_2$ and create moiré patterns of different geometries. To gain deeper insights into the atomic-scale mechanism governing the moiré structure, we acquire atomic-scale plan-view HAADF-STEM images of the sample. The reconstructed, virtual dark-field STEM images clearly show the stripe and distorted hexagon moiré domains with distinct periodicities (Fig. 3a,d). The symmetry and magnitude of the applied strain field vary with the spatial position relative to the stressor film edge and determine the moiré pattern geometry: Stripe patterns due to uniaxial strain away from the stressor edge (Fig. 3b) and distorted hexagonal moiré networks due to biaxial strain close to the stressor edge (Fig. 3e).

Atomically resolved STEM images were acquired across the domain boundary for both moiré patterns (Fig. 3c,f and Supplementary Fig. 7), which reveal changing stacking configurations. Schematics of two MoS$_2$ layers experiencing heterostrain and an interlayer shift, are drawn below the STEM images. These stacking arrangements reproduce the observed STEM images (see Supplementary Fig. 8 for simulations). Thus, a continuous change in atomic registry occurs over the span of ~7 nm-wide domain boundary that separates neighboring 2H-stacking domains. From left to right, the stacking sequence goes from 2H to the bridge (Br) region, then to saddle point (SP) stacking, back to Br, and finally to 2H stacking. Importantly, in the stripe moiré region (Fig. 3c), in-plane lattice displacements are along the zigzag direction across the domain boundary, whereas in the hexagonal moiré region (Fig. 3f) the lattice shifts are along the armchair direction. Previous studies have shown that interlayer sliding along the zigzag direction avoids unfavorable metal-metal and chalcogen-chalcogen overlap, which stabilizes stripe domains, while sliding along the armchair direction allows the lattice to relax into hexagonal networks [39]. Our observations are consistent with this microscopic understanding. The uniaxial strain drives interlayer displacements along the zigzag direction, leading to stripe moirés away from the stressor, while the biaxial strain couples to both crystallographic axes and leads to lattice shifts along the armchair direction, producing a distorted hexagonal moiré structure near the stressor edge.

**Polarization textures in moiré patterns**

The continuous stacking changes over the domain boundaries can induce polarization locally due to shifts in the atom registry. The interlayer shifts modulate charge redistribution and produce electric polarization with a spatially varying texture [11]. We employ nanobeam 4D-STEM combined with a cepstral-based algorithm of the exit wave imaginary cepstrum (EWIC) [55] to directly measure the polarization at domain boundaries. In EWIC, the logarithm of nanobeam electron diffraction (NBED) patterns at each scan position are Fourier transformed, and the imaginary component of the transformed data is analyzed. This suppresses intensity variations of NBED coming from extrinsic factors such as sample thickness, tilt, and noise, isolating asymmetries that correspond to intrinsic local polar displacements. Unlike Bragg-disk intensity methods, EWIC compiles information from the entire diffraction pattern, making it dose-efficient and resolving displacement vectors with picometer precision, enabling robust polarization mapping in beam-sensitive 2D materials. Figure 4a,e show the reconstructed ADF STEM images of the stripe and hexagonal moirés, respectively. Representative NBED patterns at the domain boundary are shown in Fig. 4b,f. The corresponding EWIC transforms (Fig. 4c,g) reveal asymmetric intensity features that can be interpreted as dipoles [56], highlighted by arrows. Each dipole encodes both the direction and magnitude of the local interlayer displacement vector, with the dipole length corresponding to the displacement magnitude. The measured dipole lengths are in the range of 1.5–2.0 Å near the domain boundaries for the stripe and hexagonal patterns, in the same range of the unit cell shifts of $a/\sqrt{3}$ = 1.8 Å along the armchair direction and of $a/2$ = 1.6 Å along the zigzag direction, where $a$ is the lattice constant, 3.16 Å. By scanning over the entire area, these local dipoles are mapped to yield the full polar displacement field of the superlattice.

The resulting polar displacement maps for the stripe and hexagonal moiré patterns are shown in Fig. 4d,h. Each pixel color encodes the direction of the local displacement vector. Within the centrosymmetric 2H domains, no polar distortion is observed. Strong in-plane polar displacements are observed in the vicinity of the domain boundaries, where the polarization vectors on opposite sides of the boundary are oriented in opposite directions. These relative displacements originate from the interlayer stacking transitions going from Br→SP→Br between adjacent 2H domains. Owing to the broken mirror symmetry, the domain boundaries develop in-plane polar textures. Hence, in-plane polarization is induced in non-polar 2H $MoS_2$ flakes due to controlled construction of moiré patterns using stressor films.

The orientation of these polarization vectors is directly coupled to the strain state and the stacking configuration. In the stripe moiré, where uniaxial strain is present, the relative sliding of adjacent layers compresses the lattice perpendicular to the boundary. This produces polarization vectors pointing toward the domain boundary under compressive strain. In the hexagonal moiré, biaxial tensile strain couples to both crystallographic axes, while an additional shear component distorts the ideal biaxial strain and skews the hexagonal domains. As a result, the polarization vectors at the boundaries are not strictly perpendicular to the domain boundaries but instead tilt, as well as outward-oriented along the hexagonal domain boundaries. These observations show that the symmetry of the applied heterostrain not only reflects the moiré geometry but also governs the orientation of the polarization vectors at the domain boundaries.

The lattice reconstruction in moiré superlattices naturally generates intralayer strain. The spatially inhomogeneous strain fields are particularly important, as they create asymmetric distributions of electron orbitals at the domain boundaries, hence giving rise to a polarization [57]. We examine the local strain distribution in the stripe and hexagonal moiré patterns and their coupling to polarization textures. Figure 5a,b show the strain maps for *xx* and *yy* components in the stripe moiré region, obtained from the cepstrum analysis of 4D STEM data. Pronounced spatial variations are observed at the moiré domain boundaries, where opposite strain values appear on either side of the boundary. This alternating strain results from lattice relaxation and interlayer coupling under uniaxial deformation. The resulting strain gradients are concentrated at the boundaries, and through flexoelectric coupling they should induce an electric polarization according to: **P** $\propto \nabla \varepsilon$ (See Supplementary Information Section 9 for details). Consistent with Fig. 4d, the corresponding strain gradient map (Fig. 5c) reveals polarization vectors pointing oppositely relative to the stripe boundaries. In the hexagonal moiré region, the strain maps (Fig. 5d,e) also show strong variations at the domain boundaries. The corresponding strain gradient map (Fig. 5f) shows a close correspondence between the strain field and polarization texture (Fig. 4h). The strain gradient orientation aligns with the polar displacement vectors and follows the distorted hexagonal boundaries. This correlation suggests that the polar distortion originates not only from relative lateral translations between layers (Fig. 4), but also from accompanying lattice distortions localized at the domain boundaries (Fig. 5).

**CONCLUSION**

In this work, we demonstrate controlled heterostrain and scalable and deterministic construction of moiré superlattice in $MoS_2$ using patterned stressor thin films and present structural information from both cross-sectional and plan-view perspectives. The applied strain is primarily transferred to the top two layers and concentrated laterally near the stressor edge. The strain field is biaxial in regions near the stressor edge and evolves into uniaxial farther away, giving rise to distinct moiré geometries: distorted hexagonal networks under combined biaxial and shear strain and stripe domains under uniaxial strain. At the atomic scale, these moiré patterns originate from in-plane lattice displacements along different crystallographic axes that relax into different types of domain networks. Stacking changes at the domain boundaries induce in-plane polar distortions, and the resulting polarization textures mirror the underlying strain state. These findings establish patterned stressor films as a scalable and controllable means to engineer designer moiré patterns, going beyond what is possible with simple twist or intrinsic lattice mismatch between layers, to induce novel properties, such as in-plane polarization in non-polar $MoS_2$.

**METHODS**

**Sample preparation**

2H-$MoS_2$ bulk crystals (HQ Graphene) were mechanically exfoliated using Scotch tape onto pre-cleaned $SiO_2$/Si substrates. Target flakes were identified with an optical microscope and transferred to $SiN_x$ membrane TEM grids and Si substrates using a polypropylene carbonate stamp. The TEM grids were mounted onto a custom holder and spin-coated with poly(methyl methacrylate) at 2000 rpm for 2 min, and baked at 170°C for 2 min. Electron-beam lithography (Nabity NPGS, Zeiss Supra SEM) was used to write 500 × 10 $\mu$m stripes spaced 10 $\mu$m apart. After pattern development, a stressor film of $Al_2O_3$ (10 nm)/$SiO_2$ (50 nm)/$Al_2O_3$ (10 nm) was deposited using e-beam evaporation at base pressure of 5e-6 torr and at a rate of ~1 Å/s. The nominal film thickness was measured using quartz crystal monitor during deposition. The film exhibits compressive stress of -105.7 MPa, Supplementary Figure 1 shows the surface profilometer data.

**Raman spectroscopy**

$MoS_2$ samples for Raman spectroscopy were exfoliated on Si substrates, followed by the same fabrication conditions for TEM samples. Raman spectra were taken using a WITec Alpha300R Confocal Raman Microscope at an excitation wavelength of 532 nm with 1800 gr/mm diffraction gratings. The laser spot size is estimated to be 0.7 μm. Raman line scans were performed along the direction perpendicular to the stressor's edge and were collected at 0.15 μm steps. The laser power was kept below 1 mW to prevent laser induced damage to the flakes. The $E_{2g}$ and $A_{1g}$ Raman modes were analyzed by fitting with Lorentzian functions to extract the frequency and full widths at half maximum.

**Scanning transmission electron microscopy characterization**

Cross-section STEM specimens were prepared using a Thermo Fisher Scientific Helios G4 UX focused ion beam (FIB) using standard lift out and thinning methods from the samples on Si substrates, with a final milling at 2 kV. Plan-view samples were prepared by transferring exfoliated $MoS_2$ flakes onto $SiN_x$ TEM grids, followed by the stressor film fabrication process described above.

Electron microscopy data were acquired using a Cs-corrected Thermo Fisher Scientific Spectra 300 X-CFEG operating at 120 kV equipped with an Electron Microscope Pixel Array Detector (EMPAD) detector for 4D-STEM. A convergence angle was 30 mrad, and inner and outer collection angles were approximately 60 and 200 mrad, respectively, for HAADF-STEM imaging. To obtain high signal-to-noise-ratio images, multiple fast acquisition images were acquired and subsequently aligned and summed using a rigid registration process optimized for noisy images. EELS was used to estimate the sample thickness of ~20 nm. The measured thickness was 0.36 mean free path t/λ, we used an inelastic mean free path of 55 nm for $MoS_2$ at 120 kV.

The 4D-STEM datasets were acquired at 120 kV with camera length of 295 mm. A 0.5-mrad convergence angle was used for the low-magnification strain maps (Fig. 2), leading to a ~3.42 nm probe size (defined by Full Width Half Maximum (FWHM) probe diameter) and ±13.4 $\mu$m depth of focus. A 3-mrad convergence angle was used for the polar distortion maps (Fig. 4,5), leading to a ~0.57 nm probe size and ±0.37 μm depth of focus. For all datasets, 2 ms dwell time and 256×256 real-space pixels were used to acquire the EMPAD 4D datasets. The data in Fig. 2 was taken at 53k magnification with a 1.88 μm field of view, and the data in Fig. 4,5 were taken at 850k magnification with a 117 nm field of view. All reconstructed virtual annular dark-field images from the 4D-STEM datasets were obtained by integrating the signal in the diffraction pattern over an angular range of 34-107 mrad at each scan position.

The EWPC method was used to measure the strain fields of $MoS_2$. The raw diffraction pattern was first transformed to logarithmic scale. The power cepstrum image was acquired through fast-Fourier transform of the diffraction. The EWPC spots corresponding to the $(10\bar{1}0)$ and $(\bar{1}100)$ spacings were measured through the center of mass method to determine the two lattice vectors along x and y directions. The strain maps for $\varepsilon_{xx}$, $\varepsilon_{yy}$, and $\varepsilon_{xy}$ components can be calculated through polar decomposition method. For EWIC analysis, the imaginary component of the cepstral transform of the logarithm of the diffraction patterns was extracted. Dipole features, appearing as positive-negative lobe pairs in the EWIC transform, were fit with a 2D Gaussian function to extract peak positions. The separation of the lobes corresponds to the magnitude of polar displacement, and their orientation defines the displacement direction. The resulting polar vector fields were mapped across the scanned region to reconstruct polarization distributions.

**Theoretical simulations**

To construct an atomistic model for $MoS_2$ multilayer, we build a five-layer $MoS_2$ structure with dimensions of approximately 158.02 × 164.22 Å. MD simulation is performed using the LAMMPS package [51]. The reactive empirical bond-order potential for $MoS_2$ (REBOMoS) [52] is employed to describe the intralayer interactions, while van der Waals forces described by the Lennard-Jones potential account for the interlayer interactions. An initial minimization step using the conjugate gradient method with convergence force criteria of $10^{-16}$ was used, which resulted in an equilibrium interlayer distance of 6.13 Å, while the intralayer Mo–Mo lattice length is found as 3.16 Å. Ramped strain is applied in the MD simulations by introducing a 0.02% compressive strain to all atoms of the top layer, such that the leftmost atoms remain fixed while the rightmost atoms are displaced by 0.02% of the top-layer length, imposing a half-symmetry boundary condition. During this process, the bottom layer and the left edge are fixed to prevent relaxation of the imposed strain. This procedure is repeated 50 times, resulting in a total compressive strain of 1% applied to the top layer. The strain transfer length scale and the Mo-Mo length is found by calculating the average strain in each layer. The average results are found by randomly selecting five sets of Mo-Mo pairs within each layer, excluding atoms at the edges to avoid boundary effects. For each pair, the interatomic distance is measured, and the corresponding strain is calculated by dividing the displacement by the original length. The results are then averaged as shown in Fig. 1g.

## ACKNOWLEDGEMENTS

The work was supported by the DOE ECMP Award number DE-SC0023905. Sample fabrication was performed in part at the Cornell NanoScale Facility, a member of the National Nanotechnology Coordinated Infrastructure (NNCI), which is supported by the National Science Foundation (Grant NNCI-2025233). Electron microscopy was supported by the National Science Foundation under Cooperative Agreement No. DMR-2039380, and additional microscopy characterization made use of the Cornell Center for Materials Research Shared Facilities, with the Thermo Fisher Helios G4 X FIB supported by NSF (DMR-1539918).

## AUTHOR CONTRIBUTIONS

Y.-M.W. and J.J.C. conceived and designed the experiments. Y.-M.W., S.D.F., and S.S. performed the STEM measurements and analysis. Y.-M.W., S.L., and N.L.W. fabricated the samples. Y.-M.W., S.L. and N.L.W. performed and analyzed Raman measurements. Y.X. and H.A. performed the theoretical simulations. Y.-M.W., S.L. and J.J.C. wrote the manuscript, with input from all authors.

## COMPETING FINANCIAL INTERESTS

The authors declare no competing financial interests.

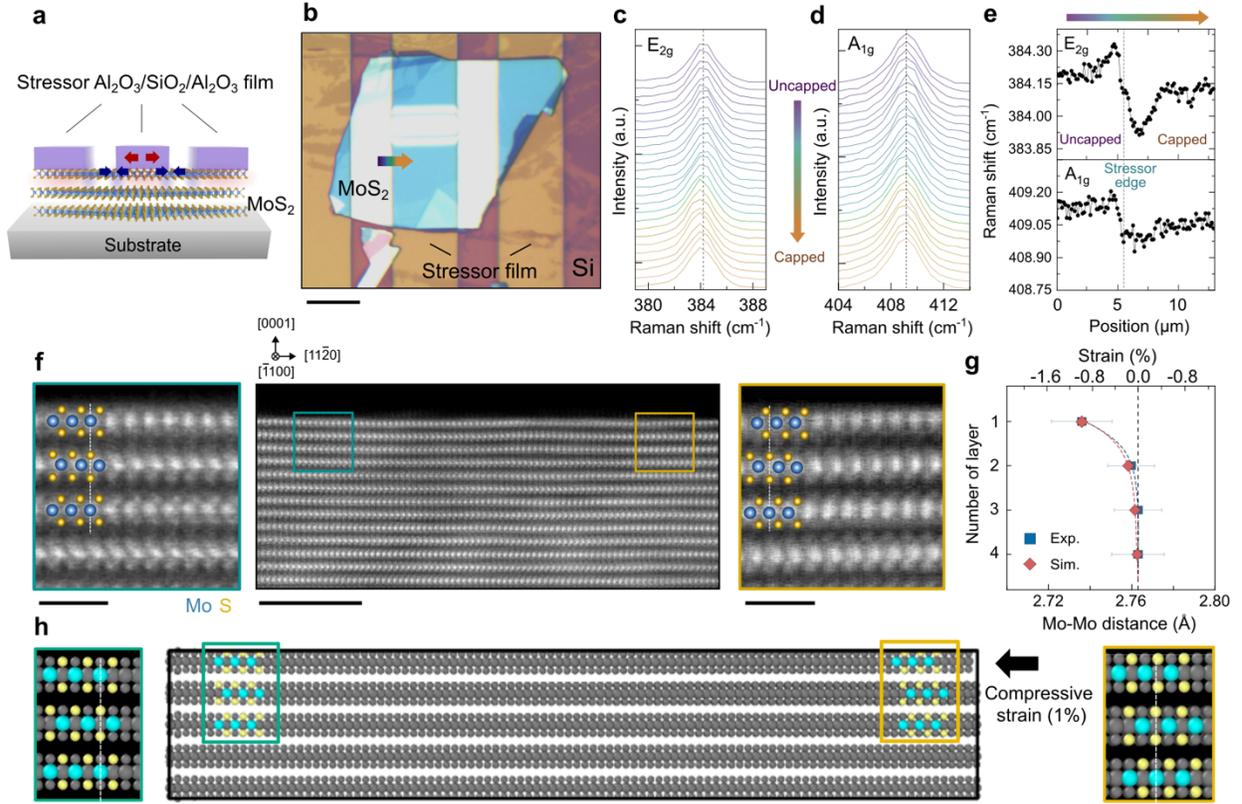

FIG. 1. **Stressor film-induced heterostrain in MoS$_2$. a,** Schematic of patterned Al$_2$O$_3$/SiO$_2$/Al$_2$O$_3$ stressor film transferring heterostrain onto MoS$_2$ locally. Red arrows indicate the strain direction. **b,** Optical image of a MoS$_2$ flake under the patterned film on the Si substrate. Scale bar, 5 $\mu$m. The Raman spectra shown in **c,d** come from regions indicated by the arrow. **c,d,** Raman spectra of $E_{2g}$ and $A_{1g}$ modes of MoS$_2$ across the uncapped to capped regions, respectively. The spectra are plotted with a vertical offset for visual clarity, with the spectra from the uncapped region at the top and from the capped region at the bottom. **e,** Corresponding $E_{2g}$ and $A_{1g}$ peak position profiles obtained by Lorentzian fitting. **f,** Cross-sectional atomic-resolution HAADF-STEM images of MoS$_2$ under the region close to the stressor edge. Scale bar, 5 nm. The panels on the left and right show magnified views of the regions highlighted by the green and yellow boxes. Scale bars, 1 nm. **g,** Layer-dependent Mo-Mo interatomic distances from experiment (blue) and simulation (red). Experimental values were obtained by quantitative atom tracking from **f**. The error bars are the standard deviations of experimental Mo-Mo distances. Simulated values were calculated from molecular dynamics for each layer. The dashed lines are polynomial fits to the data, drawn as a visual guide to the overall trend. **h,** Simulated atomic configuration model of MoS$_2$ multilayer after application of 1%

compressive strain to the top layer. All atoms are rendered using gray color to enhance the visualization of a few select atoms of Mo (cyan) and S (yellow). Magnified views of the green and yellow boxed regions are shown for comparison with experiment.

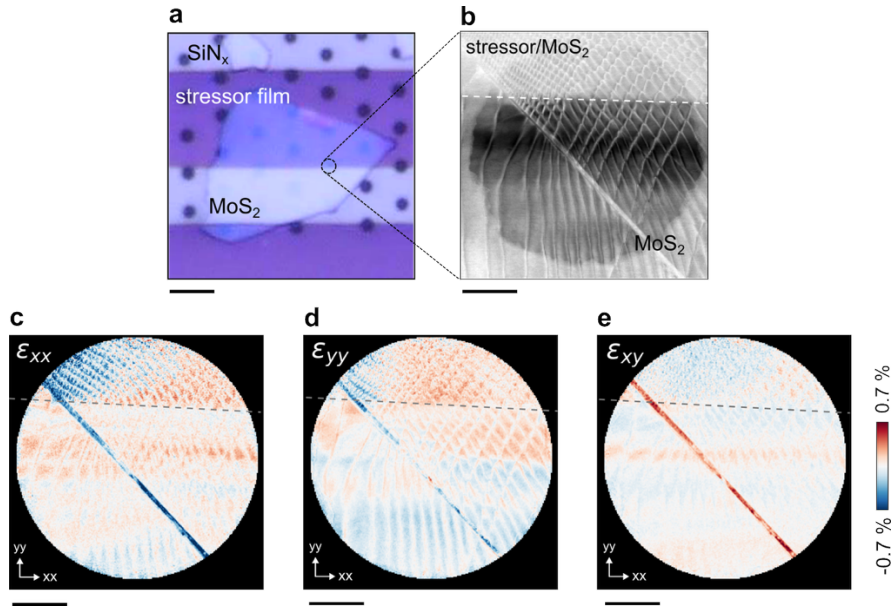

FIG. 2. **Evolution of strain distribution in MoS$_2$ under the stressor film. a,** Optical microscopy image of MoS$_2$ under the patterned stressor film. Scale bar, 5 $\mu$m. **b,** Reconstructed virtual annular dark-field image from the 4D-STEM dataset, obtained by integrating the signal in the diffraction pattern over an angular range of 34-107 mrad at each scan position. The image corresponds to a zoomed-in region near the stressor film edge marked in **a**. Scale bar, 500 nm. **c-e,** Strain maps for *xx*, *yy*, *xy* components, obtained from the same region in **b** through exit wave power cepstrum analysis. The overlaid dashed lines depict the stressor film edge. Scale bars, 500 nm.

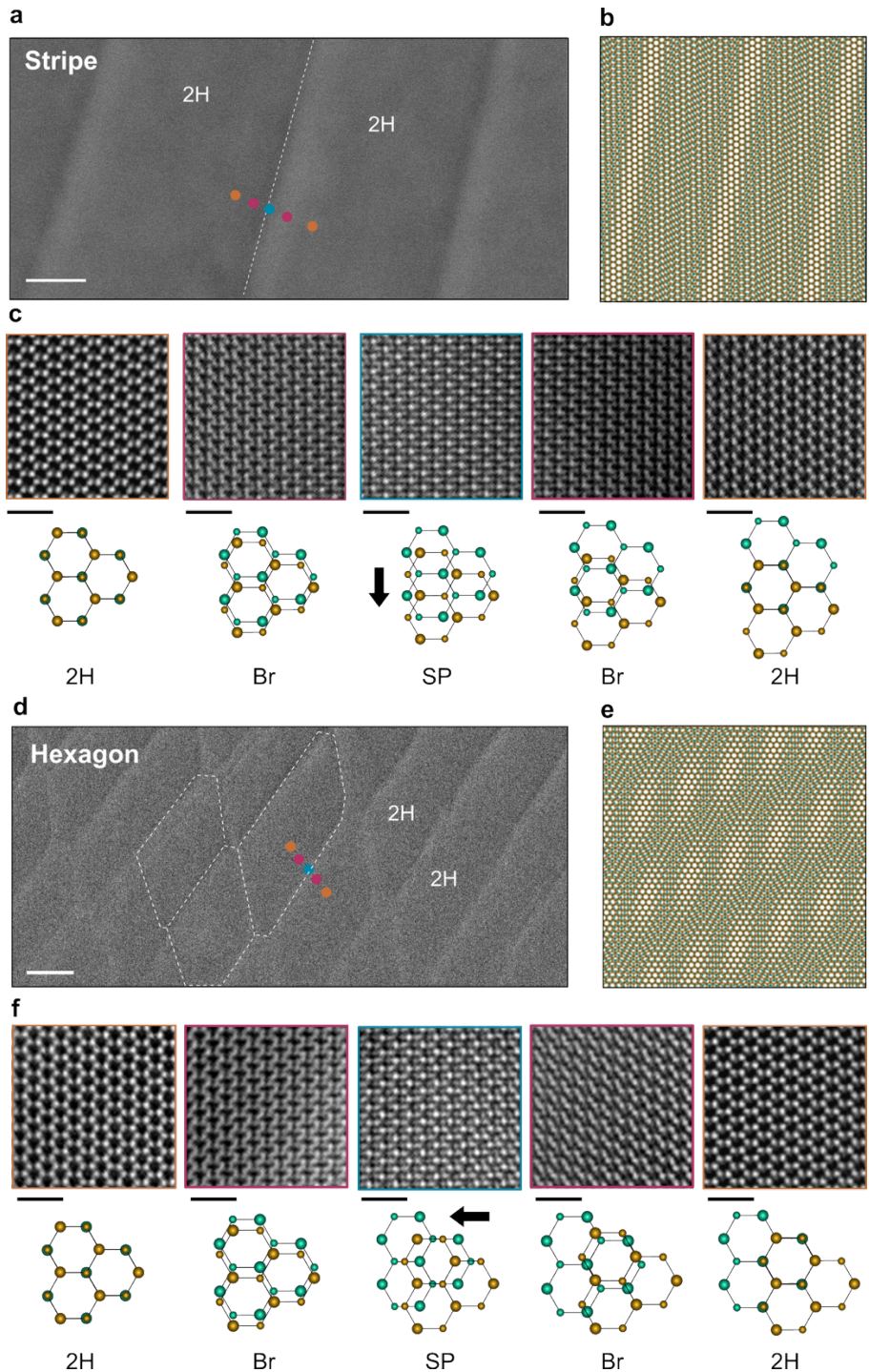

FIG. 3. **Atomic structure of moiré domains in MoS$_2$. a,** Plan-view low-magnification HAADF-STEM image of stripe-type moiré with domain boundaries highlighted by white dashed lines. Scale bar, 30 nm. **b,** Schematic of rigid moiré pattern generated by uniaxial and shear heterostrain. **c,** Zoomed-in filtered HAADF-STEM images of selected areas marked by colored dots across a domain boundary in **a** from 2H,

to the bridge region (Br), then to saddle point (SP), and back to Br, and 2H stackings with corresponding atomic schematics. The arrow indicates the lattice shift along the zigzag direction. Scale bars, 1 nm. **d,** Low-magnification HAADF-STEM image of distorted hexagonal moiré with domain boundaries highlighted by white dashed lines. Scale bar, 50 nm. **e,** Schematic of rigid moiré pattern generated by biaxial and shear heterostrain. **f,** Zoomed-in filtered HAADF-STEM images of selected areas across a domain boundary in **d** with corresponding atomic schematics. The arrow indicates the lattice shift along the armchair direction. Scale bars, 1 nm.

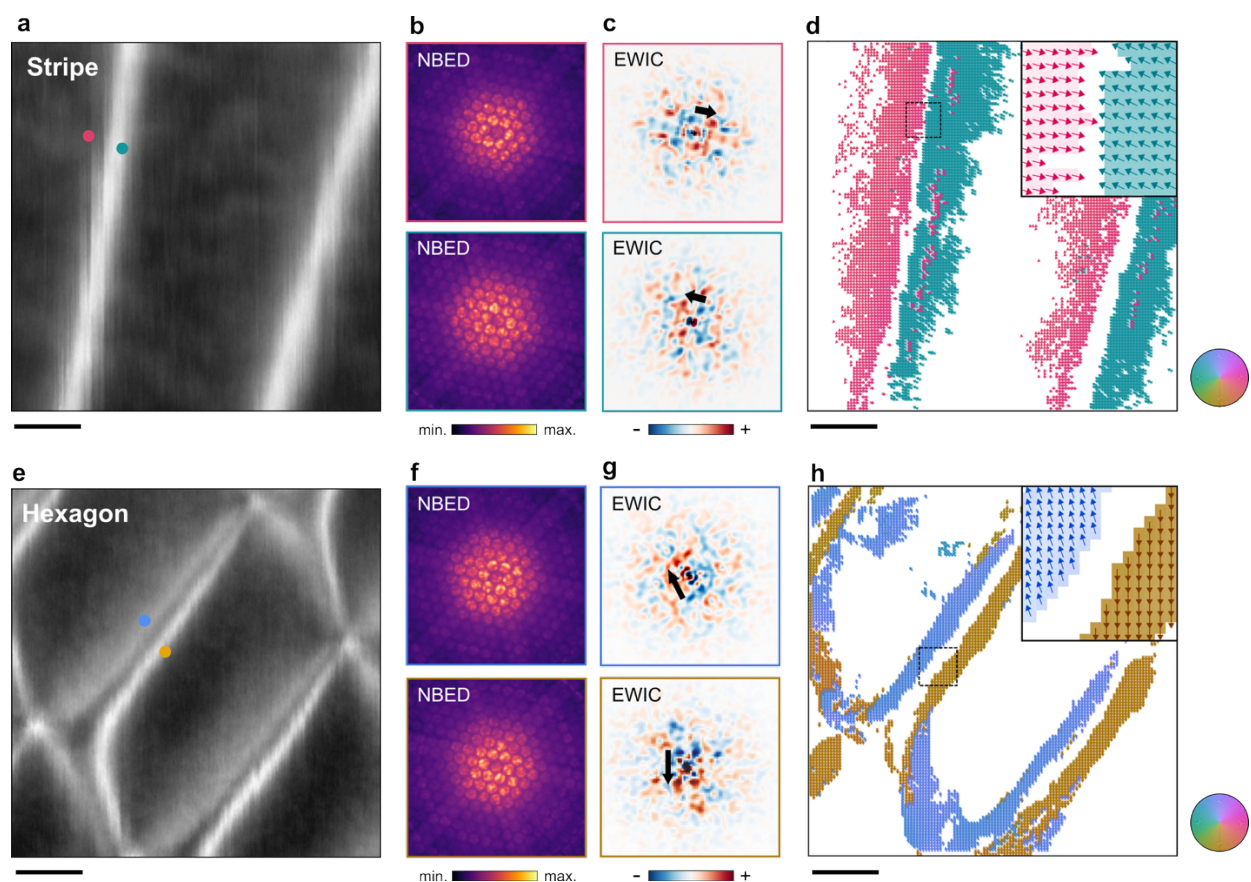

FIG. 4. **In-plane polar textures in heterostrained MoS$_2$. a,e,** Reconstructed virtual annular dark field images of stripe and hexagonal moiré patterns, respectively. **b,f,** Representative nano-beam electron diffraction (NBED) patterns with logarithmic intensity scale from the regions marked by colored dots in **a,e**. **c,g,** Corresponding exit wave imaginary cepstrum (EWIC) patterns calculated from **b,f**. The arrows indicate the directions of polar displacements. **d,h,** In-plane polarization maps of the stripe and hexagonal

moiré, respectively. Color denotes direction of polarization vectors, as specified by the color wheel, and strength gives magnitude. Insets show zoomed-in views of the maps from the dashed boxes. All scale bars, 20 nm.

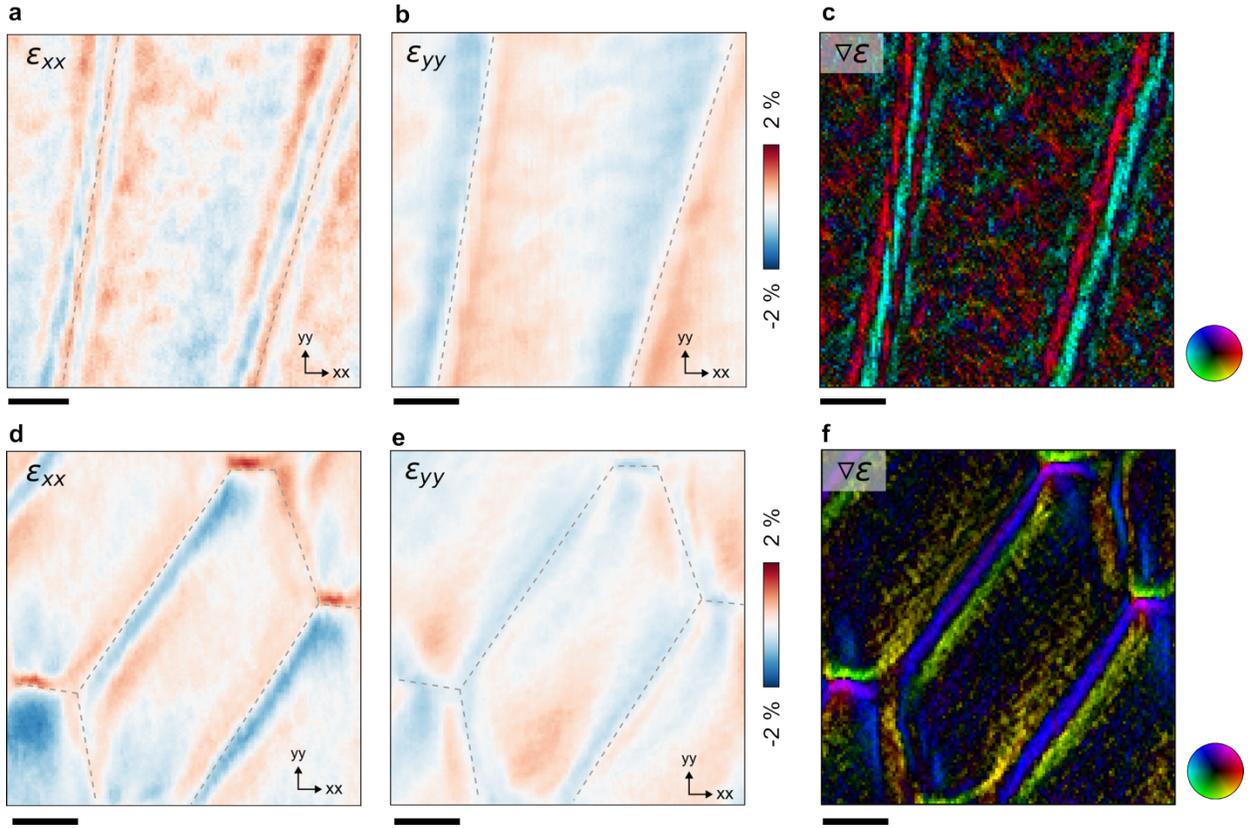

FIG. 5. **Strain-gradient induced polarization at domain boundaries. a,b,** Strain maps for *xx* and *yy* components of the moiré stripe. The overlaid dashed lines depict the moiré domain boundaries. **c,** Strain gradient map calculated from the derivative of the strain along *x* and *y* axis in **a,b**. **d,e,** Strain maps for *xx* and *yy* components of the moiré hexagon. **f,** Strain gradient map calculated from the derivative of the strain along *x* and *y* axis in **d,e**. Color denotes direction of polarization vectors and strength gives magnitude. All scale bars, 20 nm.